# Rethinking Loss of Available Work and Gouy-Stodola Theorem

Yaodong Tu and Gang Chen*

Department of Mechanical Engineering, Massachusetts Institute of Technology, Cambridge, MA 02139

*  **To whom correspondence may be addressed.** E-mail: gchen2@mit.edu

**This PDF file includes:**

    Main Text
    Figures 1 to 4

**Abstract**: Exergy represents the maximum useful work possible when a system at a specific state reaches equilibrium with the environmental dead state at temperature $T_o$. Correspondingly, the exergy difference between two states is the maximum work output when the system changes from one state to the other, assuming that during the processes, the system exchanges heat reversibly with the environment. If the process involves irreversibility, the Guoy-Stodola theorem states that the exergy destruction equals the entropy generated during the process multiplied by $T_o$. The exergy concept and the Gouy-Stodola theorem are widely used to optimize processes or systems, even when they are not directly connected to the environment. In the past, questions have been raised on if $T_o$ is the proper temperature to use in calculating the exergy destruction. Here, we start from the first and the second laws of thermodynamics to unambiguously show that the useful energy loss (*UEL*) of a system or process should equal to the entropy generation multiplied by an equivalent temperature associated with the entropy rejected out of the entire system. For many engineering systems and processes, this entropy rejection temperature can be easily calculated as the ratio of the changes of the enthalpy and entropy of the fluid stream carrying the entropy out, which we call the state-change temperature. The *UEL* is unambiguous and independent of the environmental dead state, and it should be used for system optimization rather than the exergy destruction.

**Significance Statement:** The classical exergy concept has become a standard tool in thermal system optimization. The exergy destruction evaluated by the Gouy-Stodola theorem, which equals the entropy generation $S_{gen}$ during the process multiplied by environment temperature $T_o$, is believed to be the available work loss of any thermal system. We show here that it represents only an aspirational situation when each part of a system directly communicates heat with the environment, which is not the case for most real energy systems. Starting from the two laws of thermodynamics, we demonstrate that the useful energy loss (*UEL*) of a system or process should equal to the entropy generation multiplied by an equivalent temperature associated with the entropy rejected out of the entire system, irrespective of whether the system is connected to the environment or not.

**Introduction**

Thermodynamic analyses often aim to find the maximum efficiency of an energy conversion system and identify the deviation from the maximum (1-4). The Carnot expression for efficiency, $\eta = 1 - T_c/T_h$, is considered the maximum of any heat engine working between two constant-temperature heat reservoirs at temperature $T_h$ and $T_c$ (1, 3-5). For finding the deviation from the maximum, the concept of exergy is frequently used (1-3, 6-15). For open systems, the exergy is defined as the difference of the availability $Ex$=A-A$_o$, where the availability of a fluid stream is defined as A=$H$-$T_o S$, with $H$ is enthalpy and $S$ the entropy of the system and $T_o$ is the dead state temperature when no system change can happen, i.e., the ambient temperature (16). The loss of exergy, or exergy destruction, is stated via the Gouy-Stodola theorem (17-22), $W_L = T_o S_{gen}$, where $S_{gen}$ is the entropy generation in the system or process. While the Carnot efficiency is only applicable to heat engines working between two constant temperature heat reservoirs, the exergy loss is believed to be applicable to any processes.

The exergy presents the maximum useful work possible when a system at a specific state reaches equilibrium with the environmental dead state (1, 3, 4). The exergy difference between two states, $\Delta Ex$, is then the maximum work output when the system changes from one state to the other, assuming that during the change, the system directly exchanges heat reversibly with the environment. If the process between the two states generates entropy, i.e., involves irreversibility, the exergy destruction is given by the Gouy-Stodola theorem. This exergy destruction ($W_L$) equals the exergy difference of the two states ($\Delta Ex$) minus the actual system work output ($W$), i.e., $W_L = \Delta Ex - W$. Since $Ex$ is relative to the dead state, the exergy destruction implies that the actual system is also connected with the environmental dead state. However, most of the time, real processes are not directly connected with the environment (11, 23-35). For example, for a gas turbine, neither the inlet nor the outlet is directly connected with the environment. Despite the disconnect, the Gouy-Stodola theorem is widely used to represent the loss of work potential in optimizing the process or system designs, which usually aim at reducing the exergy destruction.

In this work, we will show that the Gouy-Stodola theorem does not represent the true loss of work ability of the systems with prescribed processes, and present an alternative to the Gouy-Stodola theorem. An example of the "prescribed" process is again the gas turbine, which goes through an adiabatic expansion process, with no direct connection with the environment. We emphasize that the essence of the second law is that any entropy flows into the system, plus the entropy generated, must be carried out by heat or another fluid stream, since work does not carry entropy. By introducing a temperature ($T_{ei}$) associated the entropy flowing into the system and another flowing out ($T_{eo}$), we can express the absolute efficiency of any processes and systems in a form identical to the Carnot expression $\eta = 1 - T_{eo}/T_{ei}$, be it thermal or chemical or both. We further show that the rate of useful energy loss (*UEL*) $\dot{Y}$ is related to the entropy generation rate and $T_{eo}$ as $\dot{Y} = T_{eo}\dot{S}_{gen}$. Although the *UEL* has a similar expression to the Gouy-Stodola theorem, they are fundamentally different since *UEL* represents the actual loss of useful work of the system or process, independent of the environmental dead state. For many practical engineering systems, we show that $T_{eo}$ equals the state-change temperature $T_{sc}$, defined as the ratio of the enthalpy change (or internal energy change) to the entropy change, which can be readily calculated. We provide a few examples to show how to calculate *UEL* for systems or processes. Our approach raises fundamental questions on the usefulness of the exergy concept and the Guoy-Stodola theorem, and presents an alternative for system optimization based on *UEL*. The generalized Carnot efficiency expression also resolves the question if chemical engines such as fuel cells is subjected to the Carnot limit with a resounding yes. Of course, given the long history of thermodynamics, any claims of a completely new approach will unlikely hold. Hence, we will start with a very brief summary of some historic development related to our work and point out comprehensive textbooks and monographs on thermodynamics which we benefitted in our research. We apologize in advance for any omission of key references.

**Brief History and Questions on the Gouy-Stodola Theorem**

Modern thermodynamics started from Sadi Carnot asking how much work is available from a heat source in his famous book published in 1824 "*Reflections on the motive power of fire*" (36), in which he anticipated the efficiency of heat engine in a form that he is now honored (5), although the absolute temperature scale was not established until 1848 by Lord Kelvin (37). The development of the first law started later, in the 1840s (38) first via the independent discovery of Robert von Mayer and Thomas Joule on the mechanical and chemical energy equivalent of heat, and generalization by William Grove, Émile Clapeyron, Hermann von Helmholtz (39), and William Rankine (38).

The evolution of the exergy analysis and the Gouy-Stodola theorem is summarized in numerous reviews (13, 21, 40-42). In 1865, Clausius gave the name "entropy" and introduced the concept of "uncompensated transformations", which is equivalent to the entropy production (43). Tait used the term "availability" in his 1868 writing: "It is very desirable to have a word to express the availability for work of the heat in a given magazine" (40, 44). Maxwell used the term "available energy" in his "theory of heat" in 1871 (45). In a paper titled "*Available energy of the body and medium*" presented in 1873 (46), Gibbs provided an analytical basis for determining the available energy in a given situation. In 1889, Gouy introduced the concept of "energie utilizable" and derived its expression for closed systems (17). In 1898, Stodola independently derived the expression of "available energy" of open systems (18). Stodola's paper gave the first derivation of what we now call the Guoy-Stodola theorem. All of these works considered systems directly connected with environment via a heat engine.

From the 1930s to 1940s, through a series of papers (47, 48) and his book (15), Keenan formalized the availability expression, recasting Gibbs results in simpler and more practical terms. Keenan's expression for the irreversibility was identical to the Gouy-Stodola theorem. Keenan's work opened the door to the modern availability analysis method, although an "exergy" coined by Rant (49) gained equal acceptance. Following Keenan, in 1956 Jan Szargut developed a series of more general exergy balance equations for both non-flow and flow system (8, 50). These studies have accelerated the applications of modern exergy analysis method on the thermal system optimization (1, 2, 9, 10, 13, 14, 21, 51, 52).

Proof of the Gouy-Stodola theorem has been provided in detail in various textbooks (1, 3, 4). We use Fig. 1(a) and 1(b) to show the key idea. Consider a heat engine taking heat $Q_h$ from a heat reservoir at $T_h$ and rejecting heat $Q_c$ to a heat reservoir at $T_c$, as shown in Fig.1(a). The work output is $W = Q_h - Q_c$, and the entropy generation is the difference of entropy rejected to the reservoir at $T_c$ and entropy transferred in from the reservoir at $T_h$: $S_{gen} = \frac{Q_c}{T_c} - \frac{Q_h}{T_h}$. The maximum work $W_{max}$ happens $S_{gen} = 0$, for which case, $Q_c$ will be $Q'_c = \frac{T_c Q_h}{T_h}$ since more heat is converted into work. One can easily show that the difference between $W_{max}$ and $W$, i.e., the *UEL*,

$$Y = W_{max} - W = Q_c - Q'_c = T_c S_{gen} \qquad (1)$$

Instead of such a direct calculation of the *UEL*, Keenan conceived a two-step process as shown in Fig.1(b) in calculating the maximum work output. In the first step, a reversible heat engine is connected to $T_h$, taking heat $Q_h$ and rejecting heat to a dead environment at $T_o$. The maximum work output for this process is the exergy of $Q_h$ at $T_h$, $Ex_h = Q_h\left(1 - \frac{T_o}{T_h}\right)$. The second step is a reversible heat pump attached between the environmental at $T_o$ and the heat reservoir at $T_c$, pumping heat $Q_c$ from $T_o$ to $T_c$, consuming the minimum amount of work, i.e., the exergy between $T_c$ and $T_o$, $Ex_c = Q_c\left(1 - \frac{T_o}{T_c}\right)$. The maximum work for such a combined system is then

$$W_{max,o} = Ex_h - Ex_c = Q_h\left(1 - \frac{T_o}{T_h}\right) - Q_c\left(1 - \frac{T_o}{T_c}\right) = W + T_0 S_{gen} \qquad (2)$$

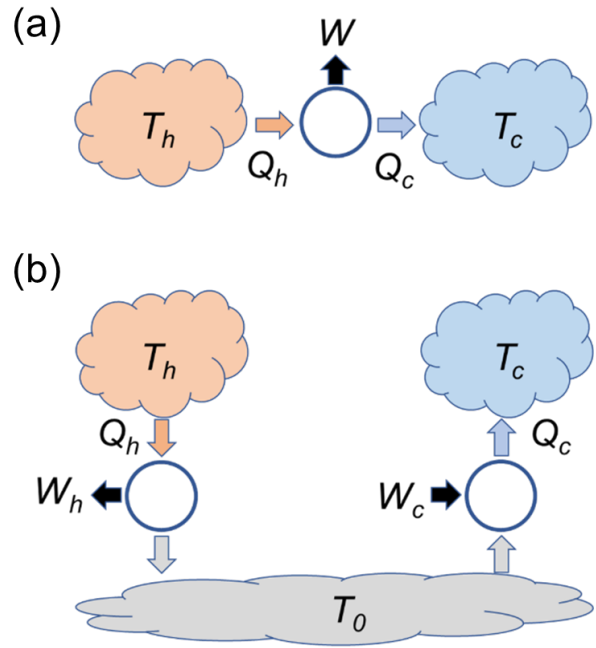

**Fig.1 Comparison between UEL and exergy destruction.** (a) A heat engine working between two thermal reservoirs, taking $Q_h$ from $T_h$ and rejecting $Q_c$ to $T_c$. For this process, the *UEL* is $T_c S_{gen}$. (b) In exergy analysis, a hypothetical two-step process is used to arrive the maximum work output possible $W_{max,o}$, with each step reversibly connected with an environmental dead-state at temperature $T_o$. The net effect also takes $Q_h$ from $T_h$ and rejects $Q_c$ to $T_c$. The difference between $W_{max,o}$ and the actual work output of real process in (a) is the exergy destruction, given by the Guoy-Stodola theorem: $W_L = T_0 S_{gen}$.

Here, we use subscript "o" for $W_{max,o}$ to denote that it is the work when the system is connected with the environment via heat engines as shown in Fig.1(b). We can see that the difference between $W_{max,o}$ and W, i.e., the exergy destruction, is $W_L = T_0 S_{gen}$. This is the Gouy-Stodola theorem. This result is rigorous under its assumption that the system outputs maximum work through reversible heat engines connected with the environment at $T_o$, both at $T_h$ and $T_c$. Although the net effect for reservoirs at $T_h$ and $T_c$ are the same—$Q_h$ is taken from $T_h$ and $Q_c$ is rejected to $T_c$—the processes in Fig.1(b) are completely different from what the originally system goes through as shown in Fig.1(a). Consequently, $W_L$ given by the Gouy-Stodola theorem does not equal to Eq. (1), which representing the *UEL* of the actual process. This discrepancy is seldomly discussed in literature. Instead, exergy analysis and the Gouy-Stodola theorem, which play important roles in most thermodynamic textbooks, are widely used in energy system optimization, regardless if the system is directly connected with the environment at both inlet and outlet. Here, by connected, we mean "entropically" connected, i.e., transferring heat via fluid flow or other means. We argue in this work that it is *UEL* that should be minimized once the process is defined, while the Guoy-Stodola theorem gives an aspirational number when the designer has complete freedom to redesign a process and makes it always accessible to the environmental dead state via entropy transfer. For the specific example in Fig.1(a), it is $T_c$, rather than $T_o$, that should be multiplied to the entropy generation to calculate the *UEL*. We note $T_c$ is the temperature at which the system rejects entropy. We will show later that the proper temperature to use in calculating *UEL* corresponds to the entropy rejection temperature.

The deficiencies of the Gouy-Stodola theorem were recognized in the past. London and his collaborator Shah (23, 24) argued that $T_o$ should be a temperature-weighting factor to be specified by the analyst based on their judgment of its relevance to the system being studied. Hesselgreaves (53) recognized the choice of the atmospheric conditions was mathematically arbitrary and proposed to use the inlet temperature of the cold fluid. Bejan (54) championed the approach of minimizing entropy generation, which is mathematically equivalent to minimization exergy destruction, but avoided the use of $T_o$. However, efficiency calculation demands converting the entropy generation back into energy units. The confusion was addressed by Lampinen and co-workers.(27, 28) In 1995, Lampinen and Keikkinen(27) defined an effective temperature to characterize the temperature of entropy transfer, and argued that this temperature should be used rather than $T_o$ in the exergy analysis. In a 2006 paper, Lampenin and Wiksten (28) further separated this effective temperature into one representing the heat transferred in (heat-absorbing temperature) and another one the heat transferred out (heat-emitting temperature), and argued that the proper temperature to use in the Guoy-Stodola theorem is the heat-emitting temperature, and in some cases, also the heat-absorbing temperature. They also used this temperature to modify the definition of exergy. This modification makes exergy no longer a state property, but depend on the actual process. The calculation of the heat-emitting and heat-absorbing temperatures could be complicated because they are proper averages of local temperatures, as in the case of heat exchangers discussed in their paper. Lampenin and Wiksten's work had been followed by a few subsequent studies.(11, 30-35) Holmberg and Laukkanen compared the Lampinen's approach to the standard exergy calculation (32), and concluded the the former gives correct results.

In this paper, we develop a theory on *UEL* and show that it can be readily calculated in a form similar to the Gouy-Stodola theorem. We argue that *UEL* and exergy destruction are two different concepts, as well illustrated in Fig.1(a) and 1(b). The exergy is always relative to an environmental dead state, and it is a state property. The exergy difference between two states represents an aspirational target: the maximum amount of work possible when both states are in communication with a dead state, typically the earth's environment. The exergy destruction as given by the Guoy-Stodola theorem hence represents the loss of work ability of an actual system relative to this aspirational maximum. The *UEL* as we will discuss below represents the actual system loss, irrelevant of the environment. This *UEL* should be the target to minimize in actual system design, once the processes of the systems are defined. Compared to Lampenin and Wiksten's theory, we focus on the consistent use of a temperature associated with the system entropy rejection in the *UEL* calculation, while the former theory sometimes had to rely on the heat emitting temperature. Our theory also allows us to calculate the entropy rejection temperature easily, without resorting to local temperature distributions, as we will show through examples.

**Useful Energy Loss (UEL) And Absolute Efficiency**

We consider a generic open system with control volume (CV), as shown in Fig. 2: fluid streams flow into the system at some inlet state(s) *in*, undergo change of states within the control volume, and exit at state(s) *out*. Work or heat transfer may occur across the boundary of the system to affect the streams' change of state or as a consequence of their change of state. We also separate heat transferred in $\dot{Q}_{in}$ and out $\dot{Q}_{out}$, since the direction of entropy flow associated with heat transfer is important. The first law of thermodynamics applied to this system reads

$$\frac{dU}{dt} = \dot{Q}_{in} - \dot{Q}_{out} - \dot{W} + \Sigma \dot{H}_{in} - \Sigma \dot{H}_{out} \qquad (3)$$

where $U$ is the internal energy of the system, $\dot{W}$ is the net rate at which work is done on the system, and $\dot{H}_{in}$ and $\dot{H}_{out}$ are the enthalpy inflows and outflows (sum is for multiple inlets and outlets), respectively. We use dot "˙" to represent rate. The second law of thermodynamics for this system is

$$\frac{dS}{dt} = \left(\int \frac{\delta \dot{Q}}{T}\right)_{in} + \left(\int \frac{\delta \dot{Q}}{T}\right)_{out} + \Sigma \dot{S}_{in} - \Sigma \dot{S}_{out} + \dot{S}_{gen} \quad (\dot{S}_{gen} \geq 0) \qquad (4)$$

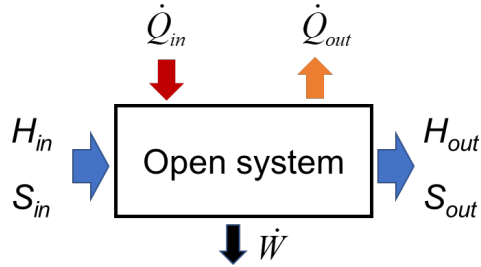

**Figure 2: An open system.** Illustration of an open system with work output $\dot{W}$, heat input $\dot{Q}_{in}$ and output $\dot{Q}_{out}$, and corresponding enthalpy and entropy flowing in and out.

where $S$ is the entropy of the CV, $\left(\int \frac{\delta \dot{Q}}{T}\right)_{in}$ and $\left(\int \frac{\delta \dot{Q}}{T}\right)_{out}$ are the rate of entropy transfer into and out of the system associated with heat transferred in and out, and $\dot{S}_{in}$ and $\dot{S}_{out}$ are the entropy inflows and outflows associated with the fluid flows, respectively. The latter two can further be related to the intrinsic entropy (s) of each stream and the mass flow rates $\dot{m}_j$, i.e., $\dot{S}_j = \dot{m}_j s_j$. Note that $\dot{S}_{j,in}$ and $\dot{S}_{j,out}$ are state properties, $\dot{S}_{gen}$ is the entropy generation rate within the CV. The entropy transfer rates $\left(\int \frac{\delta \dot{Q}}{T}\right)_{in}$ and $\left(\int \frac{\delta \dot{Q}}{T}\right)_{out}$ are process dependent, with T the temperature at the boundary where heat transfer $\delta Q$ happens. The essence of the second law is that the entropy generation rate $\dot{S}_{gen}$, which is also process dependent, is always positive and at best is zero when the process is entirely reversible. Since the entropy generation is always positive, at steady-state, the entropy transferred out, due to heat transfer $\dot{Q}_{out}$ and or carried out by the fluid $\dot{S}_{out}$ must be larger than that transferred in due to heat transfer $\dot{Q}_{in}$ and fluid flow in $\dot{S}_{in}$. Any system that outputs work must also have means to reject the entropy coming in and generated in the system, since work does not carry entropy. So, the total input energy (in rate, i.e., power unit) of the system is $(\dot{Q}_{in} + \Sigma \dot{H}_{in})$ and the total energy leaving the system is $(\dot{Q}_{out} + \Sigma \dot{H}_{out})$. The corresponding input entropy of the engine is $\left[\left(\int \frac{\delta \dot{Q}}{T}\right)_{in} + \dot{S}_{in}\right]$ and entropy rejected is $\left[\left(\int \frac{\delta \dot{Q}}{T}\right)_{out} + \dot{S}_{out}\right]$. To simplify the analysis, we define an entropy flow-in temperature, $T_{ei}$, and the entropy flow-out temperature, $T_{eo}$, as following:

$$T_{ei} = \frac{\dot{Q}_{in} + \Sigma \dot{H}_{in}}{\left(\int \frac{\delta \dot{Q}}{T}\right)_{in} + \dot{S}_{in}} \quad \text{and} \quad T_{eo} = \frac{\dot{Q}_{out} + \Sigma \dot{H}_{out}}{\left(\int \frac{\delta \dot{Q}}{T}\right)_{out} + \dot{S}_{out}} \tag{5}$$

Clearly, the temperatures thus defined are not physical ones that can be directly measured, but they are meaningful in the sense that they are associated with entropy flow. One can argue that the Kelvin temperature scale is a special case of such definitions, as it is defined thermodynamically as $T = (\partial S/\partial U)_{V,N}$ for an isolated system in equilibrium with energy U, volume V, and number of particle N. Then the combination of the first and second laws can be written as,

$$\frac{d(U - T_{eo}S)}{dt} = -\dot{W} + (\dot{Q}_{in} + \Sigma \dot{H}_{in})\left(1 - \frac{T_{eo}}{T_{ei}}\right) - T_{eo}\dot{S}_{gen} \tag{6}$$

In steady state, the power output of the system is

$$\dot{W} = (\dot{Q}_{in} + \Sigma \dot{H}_{in})\left(1 - \frac{T_{eo}}{T_{ei}}\right) - T_{eo}\dot{S}_{gen} \tag{7}$$

We can define an absolute efficiency, $\eta_a$, as the ratio of the power output and the total energy input, which reaches its maximum when the entropy generation is zero

$$\eta_a = \frac{\dot{W}}{\dot{Q}_{in}+\Sigma \dot{H}_{in}} \quad \text{and} \quad \eta_{a,max} = 1 - \frac{T_{eo}}{T_{ei}} \quad (8)$$

This expression is identical in form to the Carnot expression and represents a generalization because it is now applicable to all thermal and chemical engines converting heat or fuel into useful work. Based on our definition, $T_{eo}$ and $T_{ei}$ are always positive, and hence the efficiency will never be over one. We call this efficiency "absolute" because it is based on the input energy, relative to the value at absolute zero Kelvin. On the other hand, conventional efficiency definition uses extractable energy in the fuels, i.e., $\sum_{j=1}^{n} \dot{H}_{j,in} - \sum_{j=1}^{k} \dot{H}_{j,out}$. Although the latter definition is more useful and practical, the absolute efficiency definition is more precise and follows the principle of the second law, i.e., entropy generation must be larger than or equal to zero. The definition of absolute efficiency reflects the fact that the entropy rejected out of the system much be larger, or at best equal to, the entropy flows into the system, since irreversibility during the process increases entropy that cannot be carried out by work. Consequently, the absolute efficiency will never be over unity, as shown in Eq.(8). An analogy is the Kelvin temperature scale, which is less convenient than the temperature measures we use daily, but it represents an absolute thermodynamic concept.

Applying the above analysis and efficiency definition to fuel cells resolves the lingering debate on fuel cell efficiency. For fuel cell reactions with products entropy larger than reactants entropy (endothermic reaction), the conventional definition based on reaction heat will lead to a maximum efficiency larger than 100% (26, 55-57). Even redefining the efficiency by adding absorbed heat from environment as heat input still leads to a maximum efficiency of 100% (26, 58). Numerous researchers had argued that fuel cells are not subject to the Carnot-cycle limitation (55, 56, 59). On the contrary, Lutz (26) and Li (58) argued both heat engines and fuel cells should be subject to the same second law limitation. Using the absolute efficiency defined by Eq.(8), such difficulties are avoided, since the second law of thermodynamics dictates that any processes involving entropy changes cannot be compensated by work alone, and entropy coming into the system must be rejected!

While the absolute efficiency may be difficult to calculate because we usually do not know the properties of real materials relative to absolute zero Kelvin, Eq. (7) clearly shows that for any process that generates entropy, the rate of *UEL* can be unambiguously written as

$$\dot{Y} = T_{eo}\dot{S}_{gen} \qquad (9)$$

Although this expression looks similar to the Gouy-Stodola theorem, no reference to an environmental dead state is needed. Hence, we can calculate *UEL* in a system or process independent of the environment. The rate of *UEL* equals to the entropy generation rate times the entropy flow-out temperature, $T_{eo}$ (we can also it the entropy rejection temperature). This is consistent with the example we give related to Fig.1(a), where $T_{eo}=T_c$ since $\dot{H}_{out}$ and $\dot{S}_{out}$ in Eq. (5) are zero.

**UEL Calculation with State-Change and Entropy Transfer Temperatures**

The entropy flow-in and flow-out temperatures defined-above are based on the absolute values of the entropy and enthalpy in and out of the system, which should theoretically refer to the state at absolute zero Kelvin, where the entropy can be determined according to the third law of thermodynamics. Engineering calculations, including the efficiency definitions, often favor the differences of properties, which simplify calculations since the reference point usually does not matter when taking the difference between two different states of the same substance.

In many practical engineering scenarios, a system possibly has multi streams, together with heat transferred in $\dot{Q}_{in}$ and heat transferred out $\dot{Q}_{out}$. We usually can separate all the fluid streams in a system into two groups (1) the energy supply whose enthalpy change is negative, i.e., $(\dot{H}_{in} - \dot{H}_{out})_s > 0$ and (2) the energy sink whose enthalpy change is positive, $(\dot{H}_{out} - \dot{H}_{in})_r > 0$. The overall energy source of the system is $[\dot{Q}_{in} + \Sigma(\dot{H}_{in} - \dot{H}_{out})_s]$ and the overall energy rejection is $[\dot{Q}_{out} + \Sigma(\dot{H}_{out} - \dot{H}_{in})_r]$. Correspondingly, the overall entropy input of the system is $[(\int \frac{\delta \dot{Q}}{T})_{in} + (\Sigma \dot{S}_{in} - \Sigma \dot{S}_{out})_s]$ and the overall entropy rejection is $\left[\left(\int \frac{\delta \dot{Q}}{T}\right)_{out} + \left(\Sigma \dot{S}_{out} - \Sigma \dot{S}_{in}\right)_r\right]$. We can redefine the entropy flow-in and flow-out temperatures as

$$T_{ei} = \frac{\dot{Q}_{in} + (\Sigma \dot{H}_{in} - \Sigma \dot{H}_{out})_s}{\left(\int \frac{\delta \dot{Q}}{T}\right)_{in} + (\Sigma \dot{S}_{in} - \Sigma \dot{S}_{out})_s} \quad \text{and} \quad T_{eo} = \frac{\dot{Q}_{out} + (\Sigma \dot{H}_{out} - \Sigma \dot{H}_{in})_r}{\left(\int \frac{\delta \dot{Q}}{T}\right)_{out} + (\Sigma \dot{S}_{out} - \Sigma \dot{S}_{in})_r} \tag{10}$$

When $\dot{Q}_{out} = 0$ and for a single stream carrying entropy out, which is often the case in many engineering systems, the entropy flow-out temperature can be expressed in terms of the ratio of the differences between enthalpy and entropy at the outlet and the inlets:

$$T_{eo} = \frac{\dot{H}_{out} - \dot{H}_{in}}{\dot{S}_{out} - \dot{S}_{in}} = \frac{h_{out} - h_{in}}{s_{out} - s_{in}} = T_{sc,o} \tag{11}$$

Here, we use a special name, the state-change temperature, $T_{sc}$, to represent this special case because it depends on the properties of the inlet and outlet. This definition of course requires $(\dot{H}_{out} - \dot{H}_{in})/(\dot{S}_{out} - \dot{S}_{in}) > 0$. We can similarly define a state-change temperature for each stream. Because the definition of $T_{sc}$ involves only state properties, they are easy to calculate, in contrast to the entropy transfer temperatures. For liquids, this becomes the familiar logarithmic temperature difference between the inlet and outlet states, $T_{sc} = \frac{T_{out} - T_{in}}{\ln\left(\frac{T_{out}}{T_{in}}\right)}$.

For a closed system, we can define the state-change temperature based on the ratio of the differences between the initial and final states' internal energy and entropy,

$$T_{sc} = \frac{U_f - U_i}{S_f - S_i} \tag{12}$$

Here the subscripts $i$ and $f$ represent the initial and final states.

Using the above definitions and eliminating $\left[\dot{Q}_{out} + \left(\Sigma \dot{H}_{out} - \Sigma \dot{H}_{in}\right)_r\right]$ by multiplying the second law, i.e., Eq. (4), with $T_{eo}$ and subtracting the obtained equation by Eq. (3), we obtain

$$\dot{W} = \left[\dot{Q}_{in} + \left(\sum \dot{H}_{in} - \sum \dot{H}_{out}\right)_s\right]\left(1 - \frac{T_{eo}}{T_{ei}}\right) - T_{eo}\dot{S}_{gen} \tag{13}$$

The square bracket represents the total energy input to the system, the first term on the right-hand side represents the maximum amount of power that can be output due to thermal and/or chemical energy input. The last term is the *UEL* as defined in Eq. (9). The efficiency as typically defined in engineering is then

$$\eta = \frac{\dot{W}}{\dot{Q}_{in} + (\sum \dot{H}_{in} - \sum \dot{H}_{out})_s} = 1 - \frac{T_{eo}}{T_{ei}} - \frac{T_{eo}\dot{S}_{gen}}{\dot{Q}_{in} + (\sum \dot{H}_{in} - \sum \dot{H}_{out})_s} \tag{14}$$

with the maximum efficiency happening when $\dot{S}_{gen} = 0$, in a form identical to Carnot efficiency, $\eta = 1 - \frac{T_{eo}}{T_{ei}}$.

In the above derivation, we multiply the second law by the temperature associated with entropy flowing out of the system, consistent with the philosophy behind the second law that entropy flows in and generated within the system must be rejected. If instead, we multiply Eq. (4) by $T_o$, and then combine with the first law, we get the Gouy-Stodola theorem. Hence, we see that the major difference is that the exergy destruction chooses the environmental temperature as the reference point, while for *UEL*, $T_{eo}$ is chosen as the reference point. These two choices have different implications. When $T_o$ is chosen, it means that the system is directly connected with the environment, i.e., entropy can be exchanged with the environment, for both the energy supplying streams and energy rejecting streams. Our choice of $T_{eo}$ makes the *UEL* calculation independent of the environment. Instead, the *UEL* depends entirely on the processes that the system goes through, with $T_{eo}$ representing these fluid streams and heat transfer carrying the entropy out of the system. Lampinen and Wiksten's approach (27, 28) is also independent of the environmental dead state. In fact, the end results of our approach are often identical. However, our definition of $T_{ei}$ and $T_{eo}$ allows us consistently calculate *UEL* based on $T_{eo}$. Because they define the heat-absorbing and heat-emitting temperatures based on $Q_{in}$ and $Q_{out}$ only, for processes that do not involve $Q_{out}$, they are forced to use the heat-absorbing temperature representing entropy transferred into the system in the UEL calculation.

In addition, since the local temperature associated with heat transfer usually varies in an energy system, such as a heat exchanger, the calculation of the heat-emitting and heat-absorbing temperatures require knowledge of local temperature distributions in the system. As we will see later, for many systems, $T_{eo}$ equals $T_{sc,o}$, which can be easily calculated from the state properties at the inlet and the exit.

The above derivation can be applied to an entire system or a subsystem of the system. Although one could calculate a $T_{eo}$ for each subsystem, we emphasize that in the *UEL* calculation, we should use the $T_{eo}$ of the final entropy rejecting stream of the entire system. It is worth further clarifying that by system, we mean its subsystems are entropically connected, i.e., via heat exchange, fluid flow, or heat engines. For example, heat rejected by a refrigerator inside a house is entropically connected to the air-conditioning system of the house. In the entire house energy optimization, $T_{eo}$ should be that of external air going through air-conditioner's radiator. Coupling of two systems by work alone does not justify treating the two systems as one, i.e., each of the two systems can have their own $T_{eo}$.

Before going into examples, we point out that the state-change temperature defined above has appeared in different studies before. For example, in an exergy analysis of power plants, Woudstra (29) defined a thermodynamic equivalent temperature that corresponds to the state-change temperature, although he still used $T_o$ in exergy analysis. The logarithmic temperature had been used in discussions of entropy generation such as Bejan (54, 60), London and Shah (24). White and Shamsundar (61) used the same temperature definition in their study of the efficiency definition of heat exchangers. The combustion temperature of a fuel cell is similarly defined (26).

**Discussion**

**UEL of Engines Working Under Variable Temperature Heat Sources/Sinks and Heat Exchangers.** We consider a heat engine taking heat out from a hot stream and rejecting heat to a cold stream, as shown in Fig.3(a). Taking both the hot and the cold stream and the heat engine as the system, the first law and second law can be written as

$$W = (\dot{H}_{in,h} - \dot{H}_{out,h}) - (\dot{H}_{out,c} - \dot{H}_{in,c}) \tag{15}$$

$$\dot{S}_{gen} = \left(\dot{S}_{out,h} - \dot{S}_{in,h}\right) + \left(\dot{S}_{out,c} - \dot{S}_{in,c}\right)$$

$$= \frac{\dot{H}_{out,c} - \dot{H}_{in,c}}{T_{sc,c}} - \frac{\dot{H}_{in,h} - \dot{H}_{out,h}}{T_{sc,h}} = \frac{\dot{Q}_c}{T_{sc,c}} - \frac{\dot{Q}_h}{T_{sc,h}} \qquad (16)$$

where $T_{sc,h}$ and $T_{sc,c}$ are the state-change temperatures for the hot and the cold streams, respectively. Eliminating $\left(\dot{H}_{c,out} - \dot{H}_{c,in}\right)$, we get

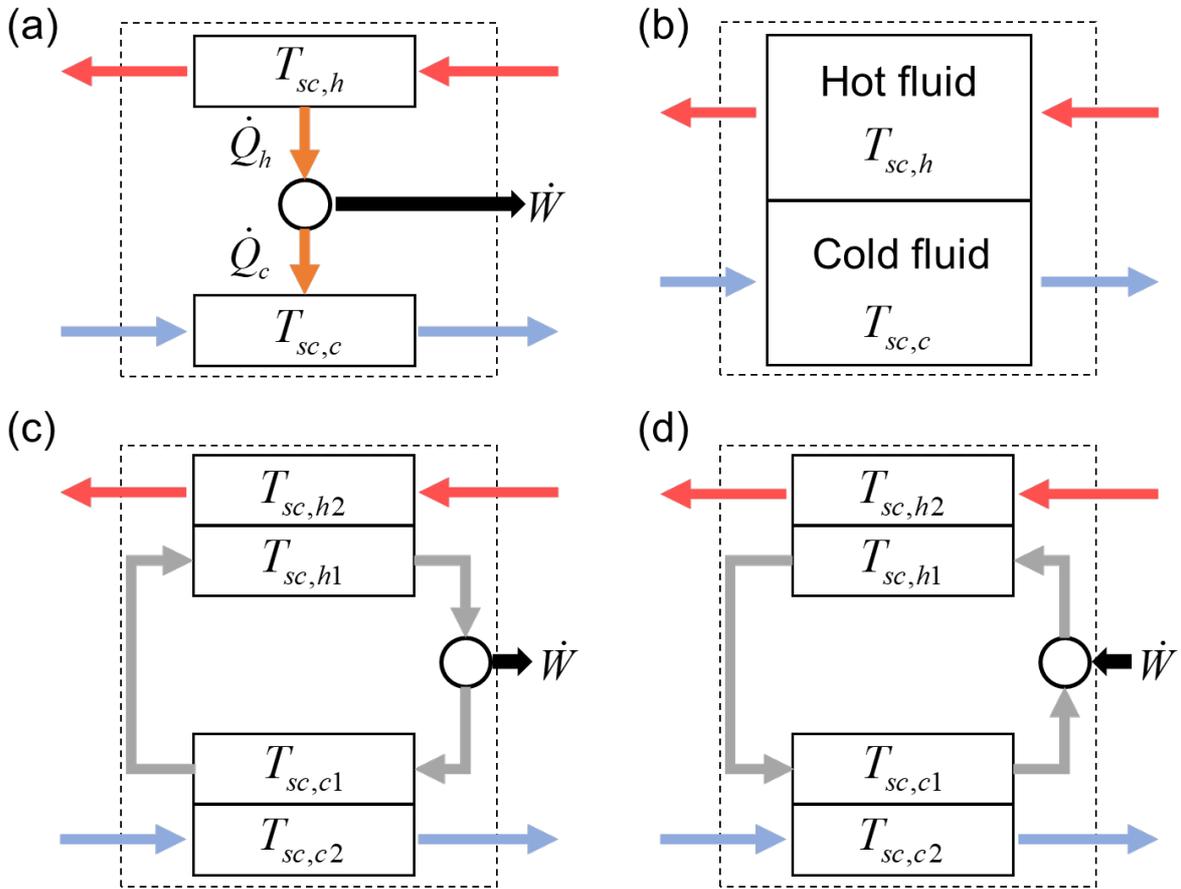

**Fig.3 Illustrations for UEL calculations for different systems:** (a) a generic heat engine, (b) a heat exchanger, (c) a power plant, and (d) an air conditioner. Cycle represents the engine, red and blue arrows represent the hot and cold streams, gray arrows represent the working fluids of a cycle, brown arrows represent the heat flow and black arrows represent the work input or output. Dashed box represents the control volume.

$$\dot{W} = (\dot{H}_{in,h} - \dot{H}_{out,h})\left(1 - \frac{T_{sc,c}}{T_{sc,h}}\right) - T_{sc,c}\dot{S}_{gen} \tag{17}$$

The above equation shows that the *UEL* of the system is $\dot{Y} = T_{sc,c}\dot{S}_{gen}$. Since the energy input to the system is $\dot{Q}_h = \dot{H}_{in,h} - \dot{H}_{out,h}$, the maximum efficiency is $\eta_{max} = 1 - \frac{T_{sc,c}}{T_{sc,h}}$. Thus, using the state-change temperatures, we can easily calculate the maximum efficiency in a form identical to that of the Carnot function under two constant temperature reservoirs and the *UEL* of the system.

Another point to make is that $\dot{Y}$ is not additive. We can take the hot fluid stream and the cold fluid stream as two different subsystems. For the hot fluid stream, $\dot{Q}_h$ goes out, and the associated entropy flow-out temperature $T_{eo,h} = \dot{Q}_h / (\int \delta\dot{Q}_h / T_h)$. The entropy generation in the hot stream is $\dot{S}_{gen,h} = (\dot{S}_{out,h} - \dot{S}_{in,h}) + \frac{\dot{Q}_h}{T_{eo,h}}$. Correspondingly, the *UEL* for the hot fluid stream is $Y_h = T_{eo,h}\dot{S}_{gen,h}$. For the cold fluid stream, the entropy flow-in temperature associated with heat transferred to the cold fluid $\dot{Q}_c = \dot{H}_{out,c} - \dot{H}_{in,c}$ is $T_{ei,c} = \dot{Q}_c / (\int \delta\dot{Q}_c / T_c)$. The entropy flow-out temperature is $T_{sc,c}$. The entropy generated in the cold stream is $\dot{S}_{gen,c} = (\dot{S}_{out,c} - \dot{S}_{in,c}) - \frac{\dot{Q}_c}{T_{ei,c}}$, and the corresponding *UEL* for the cold stream is $\dot{Y}_c = T_{sc,c}\dot{S}_{gen,c}$. Since the heat engine itself is assumed to be reversible, we have $\frac{\dot{Q}_h}{T_{eo,h}} = \frac{\dot{Q}_c}{T_{ei,c}}$. The sum of $Y_h$ and $Y_c$ is

$$Y_h + Y_c = T_{eo,h}\dot{S}_{gen,h} + T_{sc,c}\dot{S}_{gen,c} \neq T_{sc,c}(\dot{S}_{gen,c} + \dot{S}_{gen,h}) \tag{18}$$

Hence, in system optimization, one should minimize the entire system *UEL* using $T_{eo}$ of the entire system, rather than the parts.

This example confirms again that the *UEL* does not depend on defining a dead state of the environment $T_o$. Since $T_{sc,c}$ is determined by the system itself, we can use *UEL* to minimize the loss of work ability of the system directly, relevant to the designed processes. Different from London's suggestion (23, 24), which replaces $T_o$ by a value choosing by the designer, the *UEL* of a system can be calculated deterministically depending on the system parameters. This example shows that the temperature that should be used in calculating the *UEL* should be the state-change temperature of the stream where the entropy of the system is rejected.

**Heat Exchanger.** The above analysis makes it easy to write down the *UEL* for a two-stream heat exchanger as shown in Fig. 3(b), which is a special case of Fig.3(a) when W=0. For heat exchanger $\dot{Q}_h = \dot{Q}_c = \dot{Q} = \dot{H}_{in,h} - \dot{H}_{out,h} = \dot{H}_{out,c} - \dot{H}_{in,c}$. In this case,

$$\dot{Y} = T_{sc,c}\dot{S}_{gen} = \dot{Q}\left(1 - \frac{T_{sc,c}}{T_{sc,h}}\right) \tag{19}$$

Hence, to reduce the *UEL* in a heat exchanger, $\frac{T_{sc,c}}{T_{sc,h}}$ should be as close as possible. Our approach, compared to the approach of Lampinen and Wiksten (28) using local temperature along the flow path to calculate the heat-absorbing and heat-emitting temperatures, is much simpler.

**Power Plant and Air-Conditioning Systems.** Real energy conversion systems such as Rankine cycle power plant and air-conditioners involve many heat exchangers and several streams, as illustrated in Fig.3(c) and Fig.3(d). For a power plant as shown in Fig.3(c), $T_{sc,h2}$ is the state-change temperature of the fuel combustion the fluid stream, which is the heat supplied. $T_{sc,h1}$ and $T_{sc,c1}$ represent the state-change temperature of the working media such as water inside the boiler and the condenser, respectively. $T_{sc,c2}$ is the state-change temperature of the power plant coolant. Here, the *UEL* of the entire power plant is still $Y = T_{sc,c2}\dot{S}_{gen}$, where $\dot{S}_{gen}$ is the total entropy generation of the entire power plant and $T_{sc,c}$ is $T_{sc,c2}$. According to the heat exchanger discussion, both $\frac{T_{sc,h1}}{T_{sc,h2}}$ and $\frac{T_{sc,c2}}{T_{sc,c1}}$ should be as close as possible to unity to reduce *UEL* loss. If we have determined the values of $\frac{T_{sc,h1}}{T_{sc,h2}}$ and $\frac{T_{sc,c2}}{T_{sc,c1}}$ at the system optimization stage, we can optimize the geometrical structure of the heat exchanger with the help of heat transfer correlations of specific heat exchanger (54, 62, 63). We note in the *UEL* calculation of the entire power plant, it is the $T_{sc,c2}$, i.e., the entropy rejection temperature of the lowest temperature stream, that should be used, while the entropy generation is that of the entire power plant. Since $T_{sc,c2}$ is different from T₀, even though the coolant stream rejects entropy to the ambient, the UEL is different from the exergy destruction given by the Guoy-Stodola theorem.

For an air-conditioner as shown in Fig.3(d), the state-change temperature for the entropy rejecting stream $T_{sc,h2}$ should be used in the *UEL* calculation. A typical air-conditioning system specifies the indoor requirement, i.e., $\dot{Q}_c$ and $\dot{m}_{c2}$, $T_{in,c2}$, and $T_{out,c2}$, from which we can calculate the state-change temperature of the indoor air $T_{sc,c2}$. If the designer also specified the outside ambient air inlet and outlet temperatures, the outside air mass flow rate $\dot{m}_{h2}$ can then be determined as follows. First, assuming the entire system is reversible, i.e., $T_{sc,h1} = T_{sc,h2}$ and $T_{sc,c1} = T_{sc,c2}$, Eq.(16) with $\dot{Q}_h = \dot{m}_{h2}(h_{out,h2} - h_{in,h2})$ leads to the minimum $\dot{m}_{h2}$

$$(\dot{m}_{h2})_{min} = \frac{\dot{Q}_c}{h_{out,h2}-h_{in,h2}} \frac{T_{sc,h2}}{T_{sc,c2}} = \frac{s_{in,c2}-s_{out,c2}}{s_{out,h2}-s_{in,h2}} \times \dot{m}_{c2} \qquad (20)$$

The real value of $\dot{m}_{h2}$ can then be calculated from

$$\dot{m}_{h2} = \frac{\dot{Q}_c+\dot{W}}{h_{out,h2}-h_{in,h2}} = \frac{\dot{Q}_c+\dot{W}_{min}}{h_{out,h2}-h_{in,h2}} + \frac{\dot{Y}}{h_{out,h2}-h_{in,h2}} = (\dot{m}_{h,2})_{min} + \frac{\dot{S}_{gen}}{s_{out,h2}-s_{in,h2}} \qquad (21)$$

In this case, with the help of the correlations between pressure drop and flow rate (64), one can evaluate the power of the pump or fan to drive the heat source or sink streams in the heat exchanger at the high temperature end for each system configuration for which the entropy generation rate $\dot{S}_{gen}$ can be calculated. One could also use the above relation to optimize $\dot{m}_{h2}$ by minimizing *UEL*.

**Adiabatic Expansion/Compression.** For adiabatic expansion as in a turbine, we know $\dot{H}_{in} - \dot{H}_{out} \geq 0$ and $\dot{S}_{in} - \dot{S}_{out} \leq 0$, consequently, $T_{sc}$ as defined by Eq. (11) would be negative. Similarly, is the case for a compressor. We cannot use the difference in properties to define efficiency. In this case, the flow at the inlet carries the entropy in, and at the exit carries entropy out. We must use the condition at the exit to define the entropy flow-out temperature, $T_{eo}$. To define such a temperature, we first envision a reversible adiabatic expansion process for a turbine. The first and second laws written for an ideal turbine are $\dot{H}_{in} - \dot{H}_{out,r} - \dot{W}_{max} = 0$ and $\dot{S}_{in} - \dot{S}_{out,r} = 0$, where the subscript r represents the reversible state. A real turbine has friction loss and the corresponding first and second laws are written as $\dot{H}_{in} - \dot{H}_{out} - \dot{W} = 0$ and $\dot{S}_{in} - \dot{S}_{out} + \dot{S}_{gen} = 0$. The rate of *UEL* is then

$$\dot{Y} = \dot{W}_{max} - \dot{W} = \dot{H}_{out} - \dot{H}_{out,r} = T_{sc,o}\dot{S}_{gen} \tag{22}$$

where $T_{sc,o}$ is the state-change temperature of an adiabatic process, which is defined by the actual exit state and exit state when the process is reversible, i.e.,

$$T_{sc,o} = \frac{h_o - h_{o,r}}{s_o - s_{o,r}} \tag{23}$$

For ideal gas, we can easily calculate the second law efficiency $\eta = \frac{\dot{W}}{\dot{W}_{max}} = 1 - \frac{T_{sc,o}\dot{S}_{gen}}{\dot{W}_{max}}$. A similar analysis can be carried out for compressor.

We comment that the same result as Eq.(22) had been obtained before by Holmberg et al. (30) in their study of turbines. London also addressed this example, and defined an average temperature which is similar to $T_{sc,o}$, although in a differential form (23). Bejan states that the temperature to be used falls somewhere between the outlet temperature of a reversible process and the real outlet temperature in steam expansion (30). Lienhard et al. commented that Gouy-Stodola theorem is not applicable for compressor (65). We also remind that even though Eq.(22) allows the calculation of the turbines own *UEL*, when calculating the *UEL* of the entire power plant, it is the $T_{eo}$ of the coolant stream that should be used.

**Isolated System with Two Objects at Different Temperatures Approaching Equilibrium.** Next, we consider an isolated system containing two parts initially at two different temperatures $T_h$ and $T_c$. For simplicity, we assume they are made of same materials with mass $m_1$ and $m_2$ and further limit to solids or liquids for which we can write simpler expressions for entropy, with the understanding that such treatment could be readily extended to gases. With time, the two sides will approach a uniform temperature $T_f$, which can be calculated as $T_f = \frac{m_1 T_h + m_2 T_c}{m_1 + m_2}$. The entropy generation during this process is

$$S_{gen} = S_f - (S_h + S_c) = c_v \left( m_1 \ln \frac{T_f}{T_h} + m_2 \ln \frac{T_f}{T_c} \right) \tag{24}$$

where $c_v$ is the constant volume specific heat. According to the Gouy-Stodola theorem, the exergy loss is than $W_{loss} = T_o S_{gen}$. However, this exergy loss does not represent *UEL* as we show below.

We can calculate the *UEL* as follows. Imagine a reversible heat engine is placed between the two sides, which will output heat from the hot side ($Q_h$) and reject to the cold side ($Q_c$). Meanwhile, the temperature of the hot side will decrease and that of the cold side increase, until a final temperature $T_{f,r}$ is reached. This final temperature can be calculated by from the fact that no entropy is generated in such a reversible process, hence, $S_h + S_c - S_{f,r} = 0$. Substituting the entropy relations, we obtain the corresponding final temperature as $T_{f,r} = \left(T_h T_c^{\gamma}\right)^{1/(1+\gamma)}$, where $\gamma = m_2/m_1$. With this temperature, we can easily calculate the heat supplied from the hot side as $Q_h = m_1 c_v (T_h - T_{f,r})$, and heat rejected to the cold side as $Q_c = m_1 c_v (T_{f,r} - T_c)$. Hence, the maximum work output, which is also the *UEL*, is

$$Y = W_{max,1} = c_v \left[ m_1 (T_h - T_{f,r}) - m_2 (T_{f,r} - T_c) \right] \tag{25}$$

Clearly, Y given by Eq. (25) does not equal to the exergy loss $W_{loss}$ given by the Gouy-Stodola theorem. If we start from $T_{f,r}$ of the final state, and connect a reversible heat engine between the system and the environment at $T_o$, we can use the same method to calculate the maximum work output of this second process as $W_{max,2} = (m_1 + m_2) c_v (T_{f,r} - T_o) \left( 1 - \frac{T_o}{T_{sc,r}} \right)$ where $T_{sc,r}$ is the state-change temperature from $T_{f,r}$ to $T_o$. We can readily verify that $W_{max,1} + W_{max,2}$ is equal to the exergy of the two objects

$$W_{max,1} + W_{max,2} = [U_h - U_{ho} - T_o(S_h - S_o)] + [U_c - U_{co} - T_o(S_c - S_o)] \tag{26}$$

Hence, if we add the work of these two reversible steps together, the total work output is same as the exergy difference of the combined system with the environment. However, in terms of the process of two objects thermalizing with each other, the loss of available work is given by Eq. (25), not by the Gouy-Stodola theorem.

What is the appropriate entropy rejection temperature for this case? We can write Eq. (25) as

$$Y = U_f - U_{f,r} = (m_h + m_c)c_v(T_f - T_{f,r}) = \frac{T_f - T_{f,r}}{\ln\left(\frac{T_f}{T_{f,r}}\right)} S_{gen} = T_{sc,f} \dot{S}_{gen} \tag{27}$$

Hence, the corresponding entropy transfer temperature is state change temperature between $T_f$ and $T_{f,r}$, similar to the turbine case.

**Direct Evaporative Cooling.** We consider next the evaporation of pure water to a final ambient vapor $(T_o, P_{v,o})$. This process is interesting for several reasons. First, evaporation is driven by the chemical potential difference between the liquid and the vapor phase. Second, due to evaporation, the water temperature is at a lower temperature $T_{ev}$ and corresponding saturated vapor pressure $P_s$. In this case, heat from the environment is transferred to water and vapor. For simplicity, we assume that water is supplied at $(T_{ev}, P_s)$, and heat is supplied at $T_o$.

We can treat the evaporation process as the water and vapor stream receiving heat from the ambient, i.e., a heat exchanger with the hot side as the environment (Fig.4). The amount of heat transferred into the stream is $\dot{Q}_c = \dot{H}_v(T_o, p_{v,o}) - \dot{H}_w(T_{ev}) = \dot{m}[L + c_{pv}(T_0 - T_{ev})]$, where L is the latent heat and $c_{pv}$ the constant pressure specific heat of vapor. The entropy flow-in temperature is $T_o$. The vapor stream carries out the entropy transferred from the environment, with the corresponding entropy flow-out temperature equal to the state-change temperature.

$$T_{sc} = \frac{\dot{H}_w(T_{ev}) - \dot{H}_v(T_o, p_{v,o})}{\dot{S}_w(T_{ev}) - \dot{S}_v(T_o, p_{v,o})} = \frac{L + c_{pv}(T_0 - T_{ev})}{\frac{L}{T_{ev}} + c_{pv}\ln\frac{T_0}{T_{ev}} - R_v \ln\frac{P_{v,0}}{P_s}} \tag{28}$$

Using the second law, we can find that the entropy generation is $\dot{S}_{gen} = \dot{Q}_c \left( \frac{1}{T_{sc}} - \frac{1}{T_o} \right)$. According to Eq. (19), the *UEL* is

$$\dot{Y} = T_{sc} \dot{S}_{gen} \tag{29}$$

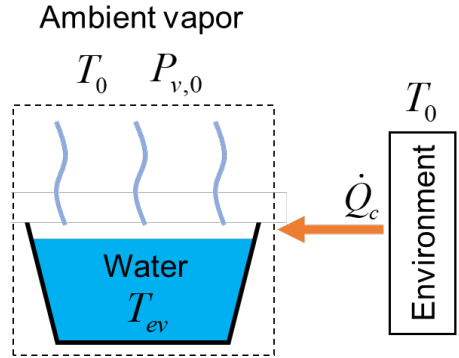

**Fig.4 Direct evaporation cooling process.** Dashed box represents the control volume (open system).

To further see that this is indeed the case, we can envision connecting a series of reversible heat engines between environment and the local vapor temperature. According to Eq. (17), for a given input $\dot{Q}_c$, the maximum work output of such a heat engine normalized to the evaporation rate is

$$\frac{\dot{W}_{max}}{\dot{m}} = \frac{\dot{Q}_c}{\dot{m}} \left( 1 - \frac{T_{sc}}{T_o} \right) = L + c_{pv}(T_0 - T_{ev}) - \frac{1}{T_0} \times \frac{[L + c_{pv}(T_0 - T_{ev})]^2}{\frac{L}{T_{ev}} + c_{pv} \ln \frac{T_0}{T_{ev}} - R_v \ln \frac{P_{v,0}}{P_s}} \tag{30}$$

One can easily show that $\dot{W}_{max}$ equals to $\dot{Y}$.

Since in this case the vapor is also connected with the environment, it is interesting to ask why Eq. (30) is not identical to the Gouy-Stodola theorem expression. The reason is that in the above case, we have fixed $\dot{Q}_c$, which is the actual heat needed to evaporate water from $T_{ev}$ to the ambient temperature. Hence, the *UEL* calculated by Eq. (30) represents the actual loss of useful work for the $\dot{Q}_c$ extracted from the environment by the vapor, i.e., for the actual process that happened. To be consistent with the exergy analysis, we can also envision a reversible heat engine between $T_o$ and $T_{sc}$, taking heat $\dot{Q}_h$ and rejection the same $\dot{Q}_c$ to the vapor stream. In this case, the heat extracted from the environmental is $\dot{Q}_h = \frac{T_o \dot{Q}_c}{T_{sc}}$. If such $\dot{Q}_h$ is actually used without output power, the entropy generation is $\dot{Q}_h \left(\frac{1}{T_{sc}} - \frac{1}{T_o}\right)$, and the correspondingly *UEL* is

$$\dot{W}_L = T_{sc}\dot{Q}_h \left(\frac{1}{T_{sc}} - \frac{1}{T_o}\right) = T_o \dot{Q}_c \left(\frac{1}{T_{sc}} - \frac{1}{T_o}\right) = T_o \dot{S}_{gen} \qquad (31)$$

This is the Gouy-Stodola theorem, where $\dot{W}_L$ is larger than $\dot{Y}$ given by Eq. (30). Clearly, $\dot{Q}_h$ is not what happened in real process. Even in this case, the Gouy-Stodola theorem does not represents the reality.

**Conclusion**

The *Gouy–Stodola theorem* for exergy destruction is derived by assuming that the system is connected to an environment reversibly. Although the theorem is rigorous for the assumptions made, the exergy destruction thus calculated does not represent the real loss of useful work for a system with already prescribed processes that are not always connected to the environment. We started from the two laws of thermodynamics and show that the *UEL* of the system can be rigorously calculated without reference to the dead state, such as the environment: $Y = T_{eo}S_{gen}$, where $T_{eo}$ is the entropy flow-out temperature of the entire system. Hence, we can state clearly the following:

*The UEL of system equals the entropy generation within the system multiplied by the entropy flow-out temperature of the system.*

We emphasize the "system", we mean the components and subsystems are entropically connected, i.e., via heat transfer or fluid flow. We show that for many engineering systems, $T_{eo}$ can be calculated from the state-change temperature of the fluid stream carrying the entropy out.

Although the *UEL* expression resembles the exergy destruction expression given by the Gouy-Stodola theorem, they are different conceptually. While the exergy represents the maximum possible work when a system reversible approaches the environment, and the exergy destruction corresponds to the loss of work relative to the maximum when the system goes through hypothetically processes entropically connected to the environment at inlets and outlets. *UEL* represents the loss of the possible work of the processes that the system actually goes through, irrespective of whether the system is connected to the environment or not. We have shown that even for systems connected to the environment such as power plants and air-conditioners, UEL does not equal to exergy destruction given by the Guou-Stodola theorem. Exergy is aspirational, pointing to the potential if a system is completely redesigned, while *UEL* is what is relevant to actual systems with prescribed processes. Engineering system optimization should target minimization of the *UEL*, not the exergy destruction.

**Acknowledgments:** G.C. would like to thank fruitful discussion with Professor Z.G. Suo, and critical comments of Professors Adrian Bejan and John Lienhard, and support of MIT. We thank Mr. Simo Pajovic, Mr. Carlos D. Diaz-Marin, and Dr. Lenan Zhang Professor John Lienhard for helpful comments on the manuscript.